# Pentazole and Ammonium Pentazolate: Crystalline Hydro-Nitrogens at High Pressure

Brad A. Steele and Ivan I. Oleynik*

*Department of Physics, University of South Florida, 4202 E. Fowler Ave., Tampa, FL 33620*

E-mail: oleynik@usf.edu

### Abstract

Two new crystalline compounds, pentazole ($N_5H$) and ammonium pentazolate $(NH_4)(N_5)$, both featuring cyclo-$N_5^-$ are discovered using first principles evolutionary search of the nitrogen-rich portion of the hydro-nitrogen binary phase diagram ($N_xH_y$, x≥y) at high pressures. Both crystals consist of the pentazolate $N_5^-$ anion and ammonium $NH_4^+$ or hydrogen $H^+$ cations. These two crystals are predicted to be thermodynamically stable at pressures above 30 GPa for $(NH_4)(N_5)$ and 50 GPa for pentazole $N_5H$. The chemical transformation of ammonium azide $(NH_4)(N_3)$ mixed with di-nitrogen ($N_2$) to ammonium pentazolate $(NH_4)(N_5)$ is predicted to become energetically favorable above 12.5 GPa. To assist in identification of newly synthesized compounds in future experiments, the Raman spectra of both crystals are calculated and mode assignments are made as a function of pressure up to 75 GPa.

# Introduction

Pentazole $N_5H$, a hydro-nitrogen molecule containing all-nitrogen aromatic ring $N_5$, has been in the focus of intensive experimental efforts of synthetic chemists for at least a century[1–4].

Although theory predicted its existence as a metastable species with a local minimum on the potential energy surface confined by a large activation barrier[5], this planar pentagon of N atoms connected to one H atom is short-lived in the gas phase, making the synthesis of this last member of azole series very difficult[6–8]. After many unsuccessful attempts, the isolated but short-lived $N_5^-$ anion has been produced in the gas phase[9,10], followed by its recent observation in a THF solution[11]. However, synthesis of hydrogen-containing $N_5H$ still remains a challenge.

To overcome this stumbling block, methods of high pressure chemistry can be employed. High pressure substantially modifies the potential energy surface thus promoting unusual chemical bonding in the condensed phase. Recently, several $N_5$-ring containing crystals have been predicted[12–14] and one of them, $CsN_5$, has been synthesized at high pressures[15]. Therefore, the difficulties of $N_5H$ synthesis can be overcome by applying high pressures to specifically chosen hydrogen and nitrogen containing precursors.

Hydronitrogens at ambient conditions display a rich variety of compounds including ammonia ($NH_3$), hydrazine ($N_2H_4$) , diimine ($N_2H_2$), triazene ($N_3H_3$), tetrazene ($N_4H_4$), hydrazoic acid or hydrogen azide ($N_3H$), and ammonium azide ($NH_4$)($N_3$)[16]. The latter compound contains an ammonium cation $NH_4^+$, which is chemically similar to heavy alkali cations[16–18]. Therefore, in an analogy with alkali pentazolates, one can also envision the existence of ammonium pentazolate ($NH_4$)($N_5$). Metallic ammonium consisting of $NH_4^+$ cations glued together by the sea of free electrons has been hypothesized to exist at high pressures inside giant planets such as Uranus and Neptune[17,18].

The high pressure chemistry of hydronitrogen systems has only recently begun to be investigated[19–26]. The experiments using the mixture of standard precursors, such as molecular $N_2$ and $H_2$, with nitrogen content ranging from 5% to 80% suggest the formation of hydronitrogen compounds as evidenced by the disappearance of $N_2/H_2$ vibrons and the simultaneous appearance of N-H stretching modes[19,20]. However, a conclusive identification of the type of new nitrogen oligomers or extended networks as well as the determination of the crystal structure have not been made. A hydronitrogen precursor ammonium azide $(NH_4)(N_3)$, containing both H and N, has not shown any signs of chemical transformation upon compression up to 70 GPa[23,27].

Theoretical calculations indicate that polymeric hydronitrogen structures can form at a relatively low pressure in a variety of H-N stoichiometries[21,24,25]. In fact, first principles calculations predict the transformation of molecular hydrogen azide to polymeric $N_3H$ at at just 6 GPa[25]. However, no such transformation was observed in experiment when hydrogen azide is compressed to high pressures[28]. Clearly there exists a gap between experiment and theory indicating that the high-pressure chemistry of hydronitrogens has not been fully explored.

The goal of this work is to study novel hydronitrogen crystalline compounds at high pressure using first principle crystal structure prediction to answer the question whether pentazolate compounds such as pentazole $N_5H$ and ammonium pentazolate $(NH_4)(N_5)$ do exist and what specific conditions are required for their synthesis. To reduce the complexity of the structure search and increase its predictive power, we specifically focus on nitrogen-rich hydronitrogen crystals, $N_xH_y$, $x \geq y$ at relatively low pressures below 60 GPa.

## Methods

The search for new nitrogen-rich hydronitrogen compounds of varying stoichiometry is performed at 30 GPa and 60 GPa by using the first principles evolutionary structure prediction method USPEX[29–31]. Two independent searches are performed with 18-24 atoms/unit cell and 8-18 atoms/unit cell to increase the probability of funding structures with nontrivial composition and geometry. The variable composition search is performed by creating numerous structures with every possible composition within the specified range of atoms in the unit cell. For instance, setting 8-18 atoms in the unit cell results in a total of 80 unique nitrogen rich compositions. Then, 300 initial crystal structures are generated, each with a composition chosen randomly from the 80 possible compositions. After the variable composition search is completed, fixed-compositions structure searches with larger number of atoms (up to 40 atoms/unit cell) are performed to find the lowest enthalpy structure for each composition.

During the structure search, the cell parameters as well as the atomic coordinates of each structure are optimized to minimize its energy, atomic forces and stresses at a given pressure using the Perdew-Burke-Ernzerhof (PBE) generalized gradient approximation (GGA)[32] within density functional theory (DFT) as implemented in VASP[33]. The PBE functional has been previously shown to give reliable results for ammonium azide[34]. For the structure search, projector augmented wave (PAW) pseudopotentials[35] and plane wave basis sets are used with an energy cutoff of 500 eV and a 0.07 Å$^{-1}$ k-point sampling. After the structure search is completed, more accurate calculations are performed to compute the convex hulls using hard PAW pseudopotentials for H and N with inner core radii of 0.423 Å for H and 0.582 Å for N, and energy cutoff of 1,400 eV and 0.05 Å$^{-1}$ k-point sampling. Charges on atoms and bond orders are calculated using LCAO code DMol[36].

The vibrational spectrum is computed using the frozen phonon method, which calculates harmonic frequencies by diagonalizing the dynamical matrix obtained from finite differencing of the atomic forces in respect to small atomic displacements. The off-resonant Raman intensities are computed 300 K and laser wavelength of 632.8 nm using derivatives atomic polarizabilities in respect to applied electric fields within a linear response framework[37].

## Results and discussion

Our structure search does find the two new crystalline materials, ammonium pentazolate $(NH_4)(N_5)$, and pentazole $N_5H$, both featuring all-nitrogen cyclic pentazoles, see Fig. 1(a,b). The first crystal, $(NH_4)(N_5)$, containing isolated ammonium cations $NH_4^+$ and pentazolate anions $N_5^-$ (Fig. 1(a)), appears on the convex hull at 30 GPa (Fig. 2). The second crystal, $N_5H$, consists of cyclo-$N_5$ covalently bonded to a H atom (Fig. 1(b)), and appears on the convex hull at 50 GPa. Both structures are predicted to be dynamically stable at 50 GPa for $N_5H$ and 30 GPa for $(NH_4)(N_5)$ as they lack any imaginary frequencies in the phonon spectrum in the entire Brillioun zone (Figures S2 and S3 respectively). The convex hull is also calculated with zero point energies and van der Waals contributions added to the formation enthalpies at 60 GPa: both $N_5H$ and $(NH_4)(N_5)$ are still on the convex hull, therefore, the entire hull does not change dramatically, see Figure S4.

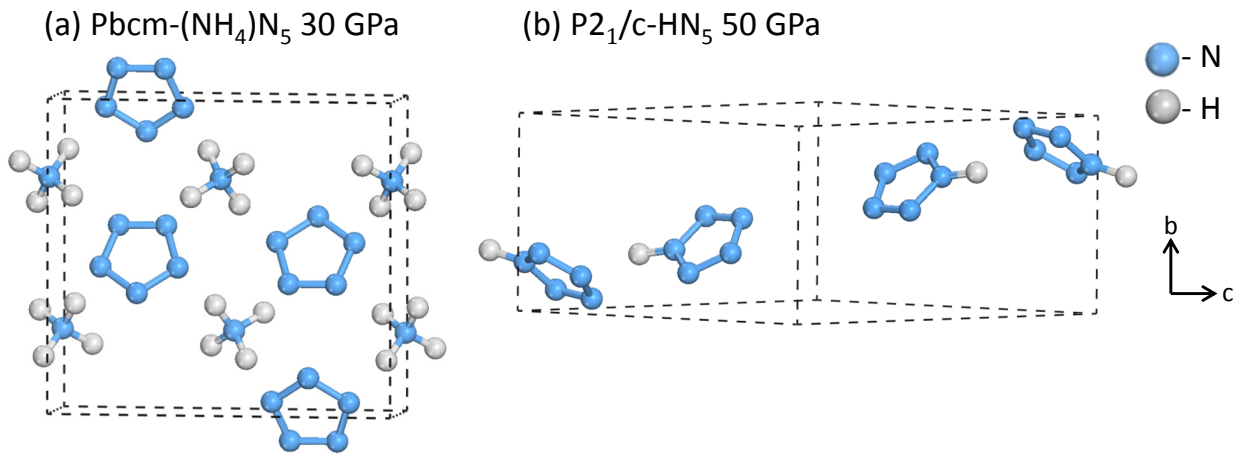

Figure 1: Crystal structures of new pentazolate compounds discovered during $N_xH$ structure search: (a) ammonium pentazolate $(NH_4)(N_5)$, and (b) pentazole $N_5H$.

Previous work by Qian $et$ $al$[25] on the structure search of hydronitrogen compounds found a plethora of novel structures, but they missed the pentazole $N_5H$ and ammonium pentazolate $(NH_4)(N_5)$ as the search was mainly conducted at very high pressures above 60 GPa. The appearance of these two new compounds dramatically modifies the convex hulls at lower pressures. In particular, the only thermodynamically stable crystals at 60 GPa are pentazole $N_5H$, ammonium pentazolate $(NH_4)(N_5)$, ammonia $NH_3$, and a crystal containing a mixture of $NH_3$ and $H_2$ molecules with net stoichiometry $NH_4$ (Fig. 2(a)). All other structures found by Qian $et$ $al$[25] with stoichiometries $N_4H$, $N_3H$, $N_9H_4$, and $N_2H$ are above the convex hull, therefore, they are metastable. The formation enthalpy of hydrazine $N_2H_4$[39] is also calculated to be above the convex hull, therefore it is also metastable. At a lower pressure of 30 GPa, ammonium pentazolate $(NH_4)(N_5)$ is still thermodynamically stable, whereas pentazole $N_5H$ is not. At 15 GPa, $(NH_4)(N_5)$ is marginally metastable, leaving only one stable compound – ammonia $(NH_3)$ (Fig. 2(a)). Another known metastable compound, ammonium azide $(NH_4)(N_3)$ with symmetry $P2/c$[40] (and stoichiometry $N_4H_4$) is also displayed at 15 and 30 GPa. Ammonium azide is predicted to undergo the phase transition to trans-tetrazene (TTZ) at 42 GPa[23], followed by another transition at higher pressure to a crystal that consists of $N_2H_5$ molecules and infinite nitrogen chains $(N_4H)$[25] therefore, the lowest enthalpy polymorph of metastable $(N_4H_4)$ with P1 space group is shown at 60 GPa as well. Several other interesting metastable higher enthalpy structures with the same stoichiometry as ammonium pentazolate $(N_6H_4)$ are discussed in the Supporting Information.

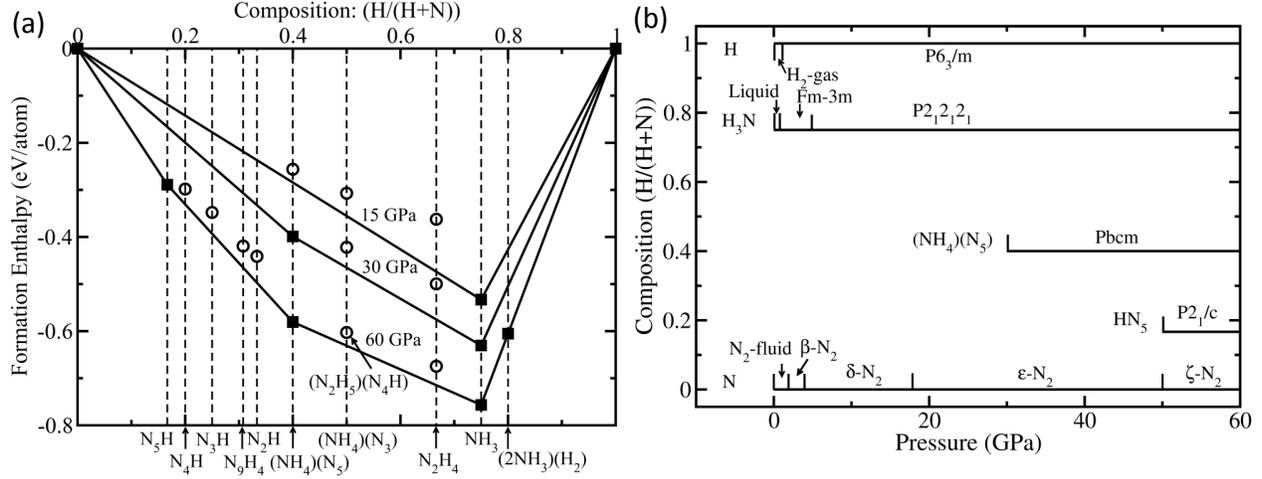

Figure 2: Hydro-nitrogen phase diagram: (a) formation enthalpy-composition convex hull, and (b) the crystal structure/pressure stability diagram. At 15 and 30 GPa the elemental N structure is $\varepsilon$-$N_2$ and at 60 GPa – cg-N. For all pressures the elemental hydrogen crystal structure has space group $P6_3/m$[38].

The cyclo-$N_5$ in both pentazole and ammonium pentazolate crystals are aromatic as (1) $4n+2$ $\pi$ electron Huckel rule is obeyed, (2) $N_5$ has planar geometry, (3) it is cyclic, and (4) every N atom has p orbital or unshared pair of p electrons able to participate in electron delocalization. The aromatic $N_5$ in these compounds has the N-N bond lengths 1.30-1.35 Å, which are intermediate between single N-N (1.45 Å) and double N=N (1.25 Å) bonds.

The strength of the aromatic N-N bond in the $N_5^-$ ring in the predicted crystals $P2_1/c$-$N_5H$ and Pbcm-$(NH_4)(N_5)$ is approximately the same as that in the gas phase. To quantify this conclusion, the total Mulliken charges, Mayer bond orders, and bond lengths in the $N_5^-$ ring in both the crystalline environment and the gas phase are calculated at 0 GPa and reported in Figure 3. In the case of $N_5H$ this comparison is straightforward as the structural unit $N_5H$ is electrically neutral. The total charge on the $N_5$ ring in the $N_5H$ crystal is 0.09 lower than that in the gas phase. This additional charge transfer between the H atom and cyclo-$N_5$ within the same structural unit is the effect of hydrogen bonding between H and N atoms belonging to different neighboring $N_5H$ molecules. These N$\cdots$H interactions result in distortion of the $N_5$ ring as well as a non-uniform charge distribution that breaks the $C_{2v}$ symmetry of the gas phase $N_5H$. In spite of these differences, the $N_5$ bond lengths and bond orders are mostly the same in both the gas phase and in the crystalline environments, except for bonds N4-N5 and N5-N1 which have a larger bond order compared to the gas phase, see Figure 3. The bond lengths of both crystalline and gas phase $N_5H$ are in a good agreement with those calculated by Ferris and Bartlett[5]. To make a unique comparison of bond orders of the $N_5$ ring in crystalline $(NH_4)(N_5)$ with those in the gas phase $N_5$, its negative charge (-0.72 e) was fixed to that in the crystalline $(NH_4)(N_5)$. The bond lengths ($\approx 1.33$ Å) and bond orders ($\approx 1.42$) in both environments are very close, the gas phase $N_5^-$ possesses five-fold $D_{5h}$ symmetry so each bond length and bond order are the same.

According to our predictions, ammonium pentazolate becomes thermodynamically stable at lower pressure (30 GPa) than pentazole $N_5H$ (50 GPa). In principle, its synthesis can be achieved by compressing the stoichiometrically balanced mixture of ammonium azide $(NH_4)(N_3)$ and diatomic nitrogen $(N_2)$ to high pressures in a diamond anvil cell to activate the chemical reaction $(NH_4)(N_3)+N_2 \rightarrow (NH_4)(N_5)$. The enthalpy difference between the products and reactants for this transformation shown in Fig. 4 displays the possibility of the phase transformation at relatively low pressure of 12.5 GPa. Ammonium azide has been reported to be chemically stable upon compression up to 70 GPa at room temperature[23].

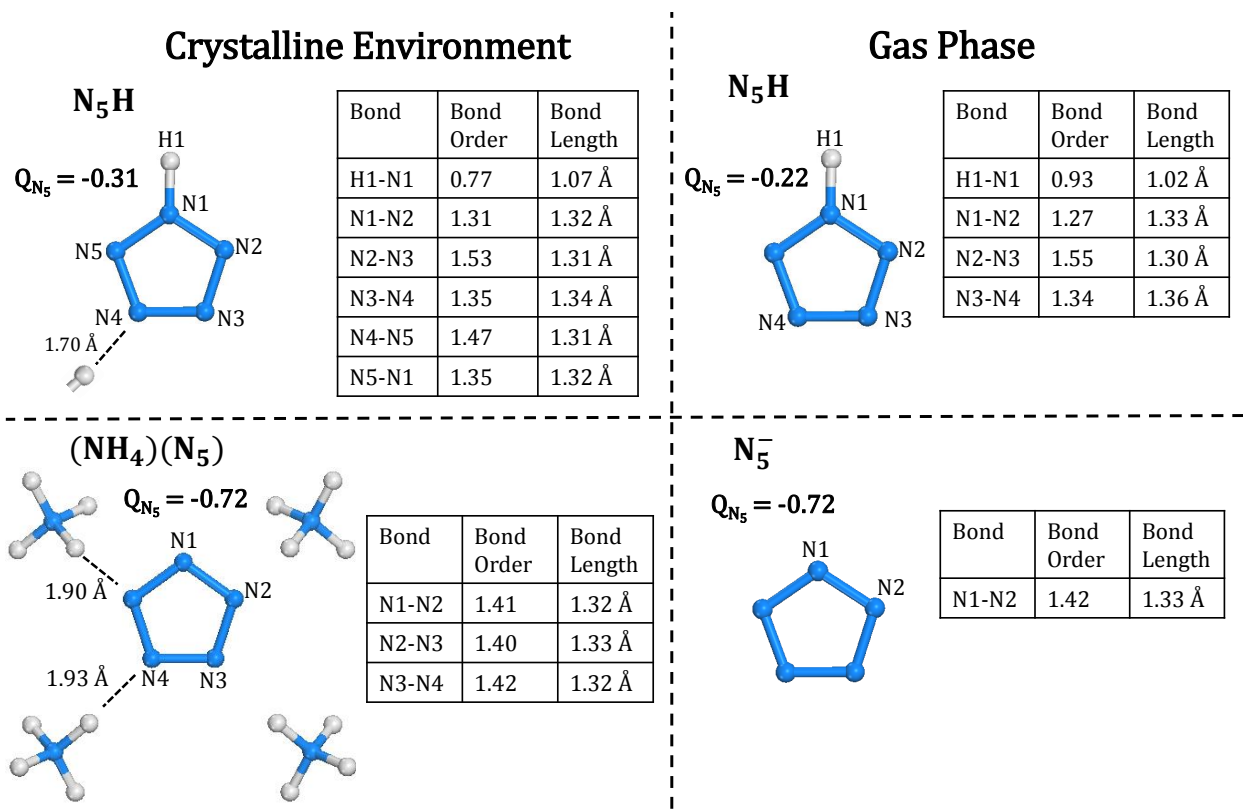

Figure 3: Total Mulliken charges, Mayer bond orders, and bond lengths in the $N_5^-$ ring in $P2_1/c$-$N_5H$ and Pbcm-$(NH_4)(N_5)$ crystals compared to those in gas-phase $N_5H$ and $N_5$. The N···H, hydrogen bonds are also shown in both crystalline environments.

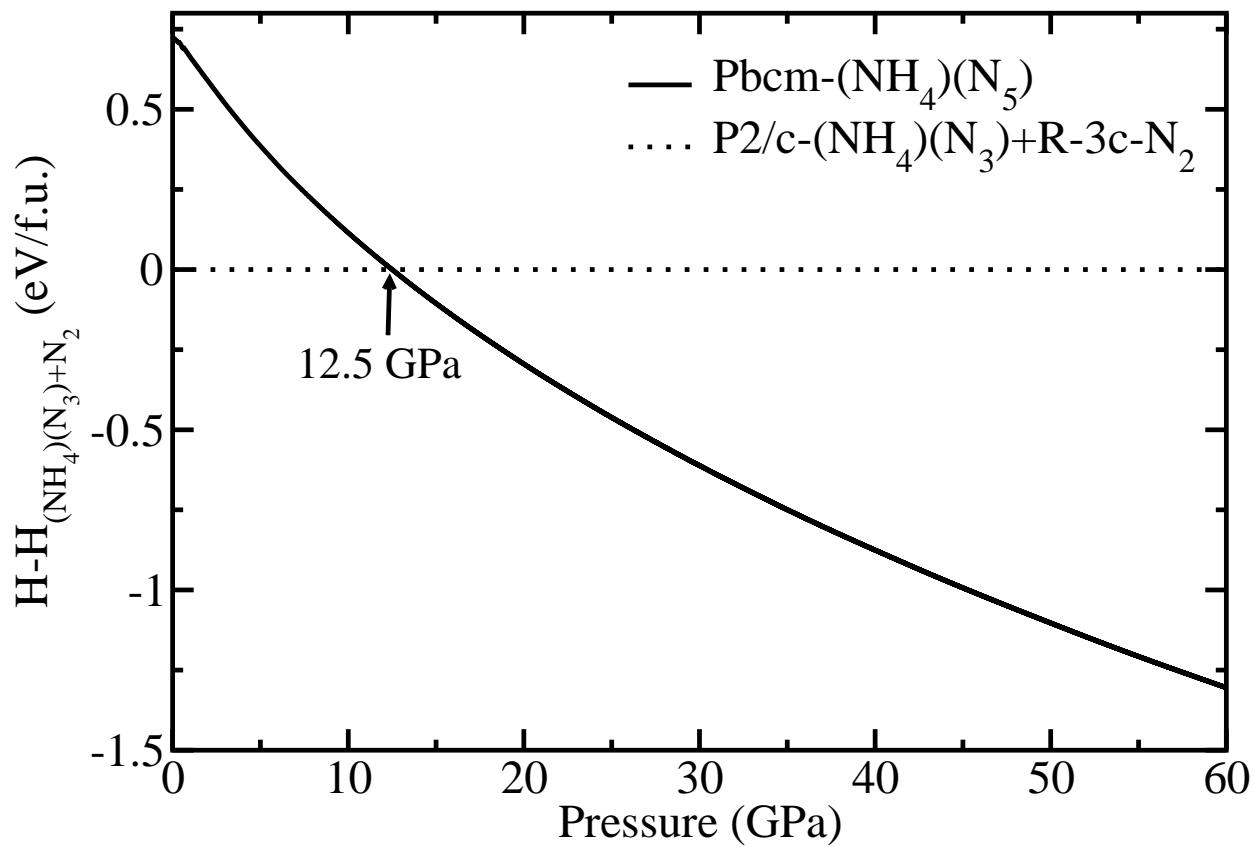

Figure 4: The enthalpy difference between the predicted ammonium pentazolate (Pbcm-$(NH_4)(N_5)$) crystal and ammonium azide (P2/c-$(NH_4)(N_3)$) plus di-nitrogen (R-3c-$N_2$). The predicted transition pressure is 12.5 GPa.

However, our calculations predict that setting up the right stoichiometry in the diamond anvil cell might activate the conversion of the the azide ($N_3^-$) anions and $N_2$ molecules to form the $N_5^-$ ring upon compression above 12.5 GPa. However, it is expected that higher pressures and temperatures might be required to overcome the significant energy barrier associated with this transformation.

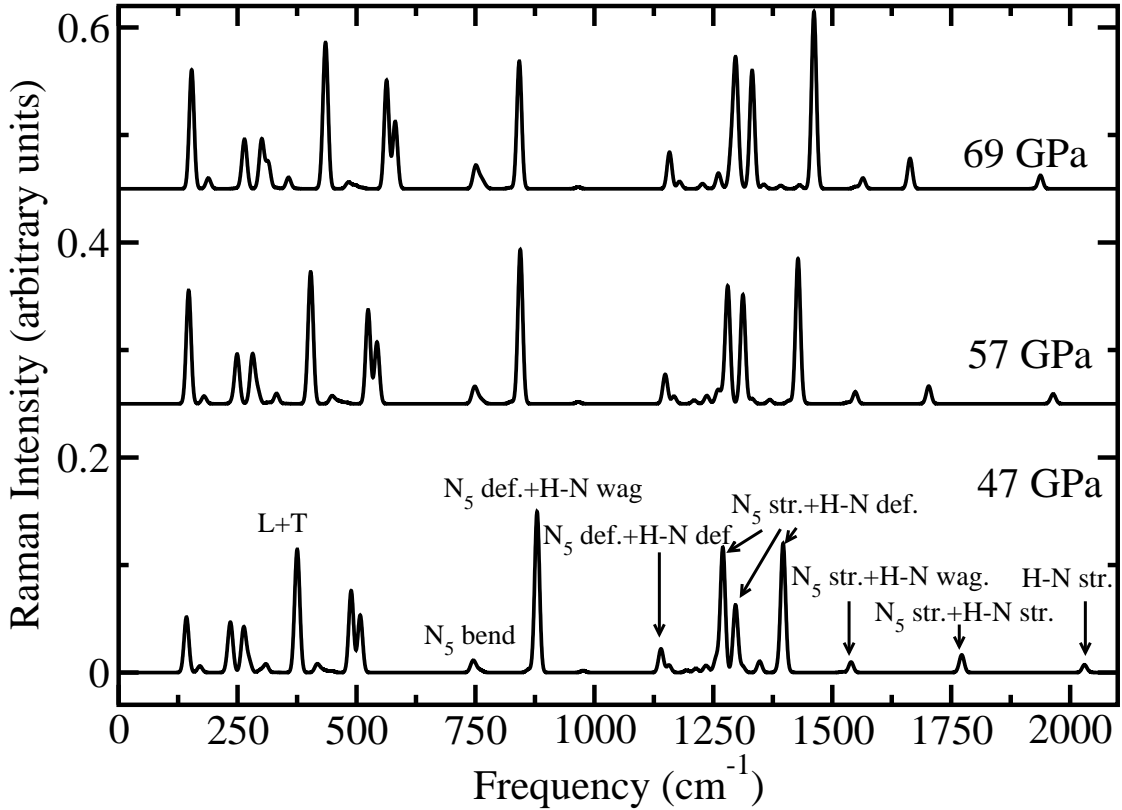

Figure 5: Calculated Raman spectra of $N_5H$ as a function of pressure; the mode assignments are given at 47 GPa.

The electronic band structure and density of states calculated at the corresponding pressures of stability, 30 GPa for $(NH_4)(N_5)$ and 50 GPa for $N_5H$ (Supporting information Figures S5 and S6 respectively), demonstrate that both materials are wide-band gap semiconductors with the band gaps 3.5 eV and 3.9 eV for $N_5H$ and $(NH_4)(N_5)$ respectively. As the band gaps of these materials are larger than the energy of visible photons ( < 3.1 eV), these materials are expected to be optically transparent at these pressures.

In order to assist in the detection of the new compounds during their experimental synthesis, the Raman spectra are calculated for both crystals as a function of pressure, see Figures 5 for $N_5H$ and 6 for $(NH_4)(N_5)$ respectively. . Mode assignments, shown in Figures 5 and 6 are made by visualizing the atomic displacements of the corresponding eigenmodes. At pressures around 50 GPa, there are several lattice (T) and librational modes (L) in the low frequency interval 0-500 cm$^{-1}$ for both crystals. At 28.6 GPa ammonium pentazolate has a few closely spaced $NH_4$ librational modes near 600 cm$^{-1}$ with a weak Raman intensity that results in one visible peak in the spectrum. Upon increase of pressure, the intensity of these modes grows while their frequencies blue shift as it is clearly seen near 745 cm$^{-1}$ at 75.5 GPa (Figure 6).

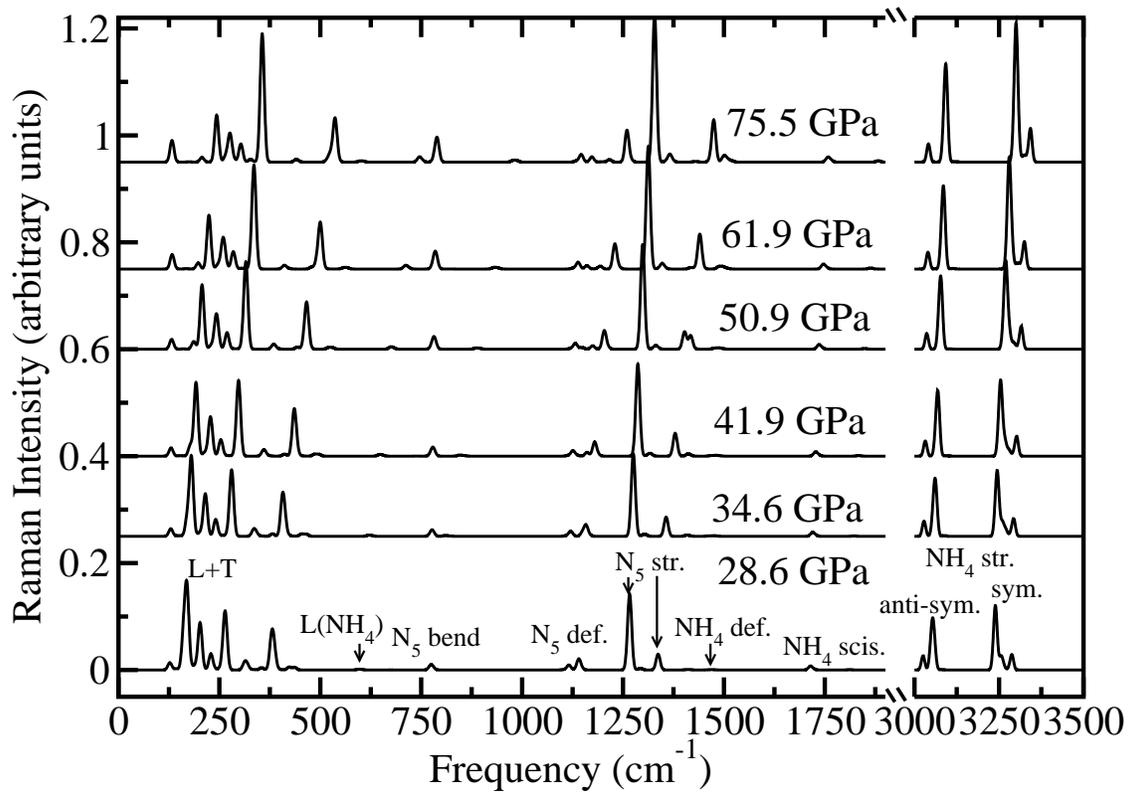

Figure 6: Calculated Raman spectra of $(NH_4)(N_5)$ as a function of pressure, the mode assignments are given at 28.6 GPa.

Pentazole and ammonium pentazolate have a weak Raman active $N_5$ bending mode around 750-790 cm$^{-1}$ with a frequency that does not shift much with pressure. Pentazole has a mode with a frequency of about 880 cm$^{-1}$ at pressures near 47 GPa with a strong Raman intensity that involves a wag or bend of the H-N bond as well as a small deformation of the $N_5$ molecule. This mode red-shifts with pressure due to a lengthening of the H-N bond upon compression. The H-N bond length is the same for each molecule in the unit cell. The H-N bond increases with pressure due to an enhanced interaction between $H^+$ and adjacent $N_5^-$ molecule. This also causes the H-N stretching mode at 2,030 cm$^{-1}$ at 47 GPa to also red shift with pressure, see Figure 5. In contrast, the H-N bond length in ammonium pentazolate decreases with pressure.

Near 1,100 cm$^{-1}$ both $N_5H$ and $(NH_4)(N_5)$ have a $N_5$ deformational mode; however, in case of $N_5H$ this mode also involves a slight H-N deformation (wag and stretch). There are additional strong peaks in the Raman spectra for $N_5H$ from these types of modes near 1,250 cm$^{-1}$ at 47 GPa. At about 50 GPa, the $N_5$ stretching (str.) mode in ammonium pentazolate has a frequency of about 1,298 cm$^{-1}$ while this mode in pentazole $N_5H$ has a larger frequency – 1,396 cm$^{-1}$. The stretching mode is different for $N_5H$ due to the slight deformation of the H-N bond in concert with the $N_5$ stretch. Pentazole has two additional modes in a frequency interval before the H-N stretch at 2,030 cm$^{-1}$, which involve an $N_5$ stretch and either a slight H-N wag or a H-N stretch at frequencies of 1,539 cm$^{-1}$ and 1,772 cm$^{-1}$ respectively, see the 47 GPa panel in Figure 5. Although these modes as well as the N$_5$ stretch in N$_5$H also involve an H-N deformation, the amplitude of the H-N deformation is small so they blue shift with pressure.

Ammonium pentazolate also has two Raman active modes that have appreciable intensity with frequencies lower than those corresponding to the H-N stretch in NH$_4$. These modes are a deformation of the NH$_4$ molecule or a scissoring (scis.) of the NH$_4$ molecule with frequencies at 28 GPa of 1,413 cm$^{-1}$ and 1,714 cm$^{-1}$ respectively, see Figure 6. There are several H-N symmetric and anti-symmetric stretching modes in the range of frequencies 3,000-3,350 cm$^{-1}$, but as can be seen from Figure 6 there are only two strong intensity bands with frequencies 3,055 cm$^{-1}$ and 3,240 cm$^{-1}$ (at 28.6 GPa) from the anti-symmetric and symmetric stretching modes of NH$_4$ respectively. Each mode blue shifts slightly with pressure similar to what is measured by Crowhurst *et al* at high pressures for the NH$_4$ stretching modes of ammonium azide[23]. The calculated relative intensities of the symmetric and anti-symmetric modes are approximately the same which differs from the NH$_4$ stretching modes measured for ammonium azide at high pressures[23], where the symmetric mode is much more intense than any other NH$_4$ stretching modes. The calculated frequencies are also slightly higher than those measured by Crowhurst *et al.* These differences in the frequency and intensity of the H-N stretching modes is likely due to the different crystalline environment of ammonium pentazolate.

## Conclusions

In summary, new hydronitrogen crystalline compounds, pentazole (N$_5$H) and ammonium pentazolate (NH$_4$)(N$_5$), have been predicted to exist at high pressures. Both consist of cyclo-N$_5^-$ anions and either ammonium (NH$_4^+$) or hydrogen (H$^+$) cations for (NH$_4$)(N$_5$) and N$_5$H respectively. First-principles evolutionary crystal structure search predicts the ther-modynamical stability of $(NH_4)(N_5)$ above 30 GPa and $N_5H$ – above 50 GPa, thus making previously proposed hydronitrogen solids consisting of hydrogen-capped long nitrogen chains unstable at these pressures. The $N_5^-$ ring in both crystals is aromatic (with bond lengths between 1.31 and 1.34 Å) that enhances the stability of this molecule. Charge transfer from the ammonium $NH_4$ to $N_5$ further enhances the stability of ammonium pentazolate reducing the pressure where it becomes stable. The chemical transformation from ammonium azide plus di-nitrogen $(NH_4)(N_3)+N_2$ to ammonium pentazolate $(NH_4)(N_5)$ is predicted to occur at pressures above 12.5 GPa. To assist in future experimental synthesis of these materials at high pressures, the pressure-dependent Raman spectra are calculated and mode assignments are made. This work demonstrates the power of first-principles structure search methods in discovering unknown crystalline materials at high pressures, thus inspiring future high-pressure experiments.

# Acknowledgement

This research is supported by Defense Threat Reduction Agency, grant HDTRA1-12-1-0023 and ARL-USF Cooperative Agreement W911NF-16-2-0022. Simulations were performed using the NSF XSEDE facilities (grant No. TG-MCA08X040), DOE BNL CFN computational user facility, and USF Research Computing Cluster supported by NSF (grant No. CHE-1531590) .

# Supporting Information Available

Details of structure prediction calculations, snapshots of metastable higher energy structures revealed during structure search, enthalpy differences, and band structure calculations. This material is available free of charge via the Internet at `http://pubs.acs.org/`.

**Corresponding Authors**

email: oleynik@usf.edu.

**Notes**

The authors declare no competing financial interests.